\documentclass{iaus}
\usepackage{graphicx}

\title[Dust] %% give here short title %%
{The Millennium Galaxy Catalogue: The severe attenuation of bulge flux by dusty spiral discs}

\author[S.P.Driver]   %% give here short author list %%
{Simon P. Driver$^1$%
 \and the MGC Team}

\affiliation{$^1$School of Physics and Astronomy, University of St Andrews, North Haugh, St Andrews, KY16 8XB, Scotland
\break email: spd3@st-and.ac.uk\\[\affilskip]}

\pubyear{2007}
\volume{IAU245}  %% insert here IAU Symposium No.
\pagerange{???--???}
\date{?? and in revised form ??}
\jname{Formation and Evolution of Galaxy Bulges}
\editors{M. Bureau, E. Anthanassoula \& B. Barbury, eds.}
\begin{document}

\maketitle

\begin{abstract}
Using the Millennium Galaxy Catalogue we quantify the dependency of
the disc and bulge luminosity functions on galaxy inclination. Using a
contemporary dust model we show that our results are consistent with
galaxy discs being optically thick in their central regions
($\tau_B^f=3.8\pm0.7$). As a consequence the measured $B$-band fluxes
of bulges can be severely attenuated by 50\% to 95\% depending on disc
inclination. We argue that a galaxy's optical appearance can be
radically transformed by simply removing the dust, e.g. during cluster
infall, with mid-type galaxies becoming earlier, redder, and more
luminous. Finally we derive the mean photon escape fraction from the
integrated galaxy population over the $0.1 \mu$m to $2.1 \mu$m range,
and use this to show that the energy of starlight absorbed by dust (in
our model) is in close agreement with the total far-IR emission.
\keywords{galaxies: bulges, galaxies: evolution, galaxies: formation,
galaxies:fundamental parameters, galaxies:luminosity function,mass
function}
\end{abstract}

\firstsection % if your document starts with a section,
\section{Introduction}
The issue of dust in galaxies has a long and heated history
(\cite{valentijn}, \cite{burstein}, \cite{disney}).  To summarise:
Selection effects coupled with simplistic slab models played havoc
with the interpretation of the available, but highly biased, galaxy
samples. Ultimately it was agreed that one cannot use the available
data to simultaneously constrain opacity and dust geometry and no
clear consensus on opacity was reached. Certainly, from within the MW,
one can see that our galaxy is optically thin away from the Galactic
plane (we can see external galaxies), and is optically thick within it
(the Galactic Centre is opague). However how the central region of our
Galaxy would appear if viewed face-on is unclear. Observations of
nearby spiral galaxies with a second spiral galaxy behind (e.g.,
\cite{keel}) suggest that, at least in the inter-arm region, galaxies
are optically thin. This is corroborated by the number-counts of
galaxies in inter-arm regions (\cite{counts}). However this does not
prove that galaxies are optically thin throughout (one can only reach
that conclusion if one adopts a simplistic slab like model, e.g.,
\cite{disney}; \cite{byun}). Recently Driver et al. (2007) use a
relatively new method to constrain the central face-on opacity of
galaxies. This involves measuring the galaxy luminosity function (GLF)
for face-on, inclined, and edge-on systems (see Fig.~1). In the
absence of dust the measured GLFs should be identical. If galaxies are
optically thick one expects edge-on systems to be severely attenuated
and the measured GLF shifted to fainter magnitudes. The larger the
shift the higher the central face-on opacity.

\begin{figure}[h]
\vspace{-5.0cm}
\includegraphics[width=\textwidth]{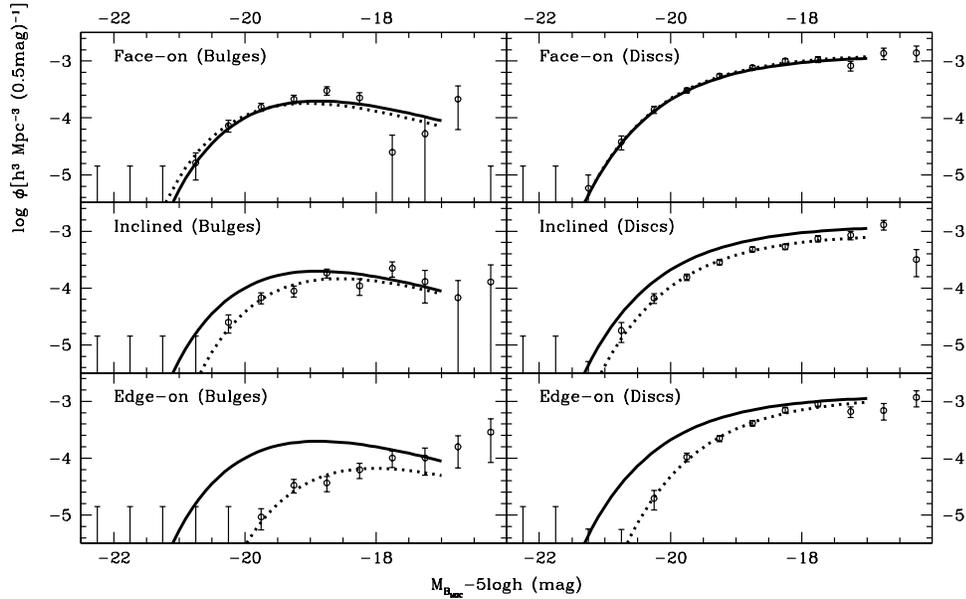}
\caption{The galaxy luminosity function of bulges (left) and discs (right) for face-on (upper), inclined (centre), and edge-on (lower) systems.}
\vspace{-0.5cm}
\end{figure}

\section*{The attenuation-inclination relation}
Fig.~1 shows GLFs measured for bulges (left) and discs (right) drawn
from face-on (top), inclined (centre), and edge-on systems
(lower). Fig.~1 shows a significant trend of greater attenuation
towards higher inclination. Fig.~2 summarises the
attenuation-inclination relation for bulges and discs across the full
inclination range (see \cite{driver07} for full details).  The data
shown in Fig. ~1 ~\&~2 has been derived from the Millennium Galaxy
Catalogue (\cite{liske}) which contains 10,095 galaxies with full
bulge-disc decompositions (\cite{allen}), and which has been shown to
be robust to surface brightness selection bias to $M_{B} < -16$ mag
(see \cite{driver05}).

\begin{figure}[h]
\vspace{-6.5cm}
\includegraphics[width=\textwidth]{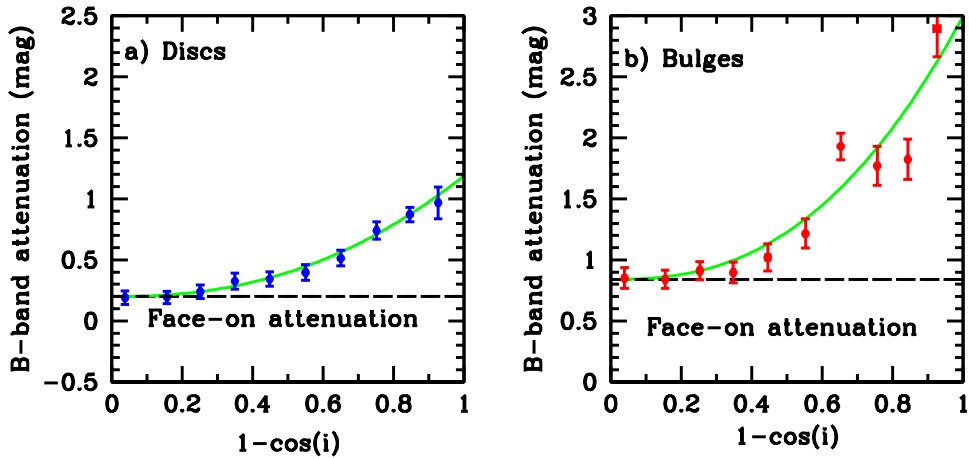}
\caption{The empirical variation of the GLR turnover point for discs (left) and bulges (right) as a function of galaxy inclination.  The face-on attenuation is model dependent, see text.}
\end{figure}

To model the attenuation-inclination relation and derive the remaining
face-on attenuation, we adopt the model of Popescu et al. (2000) and
Tuffs et al. (2004) (see also \cite{moll}). This dust model
incorporates three distinct components:

\noindent
1. An optically thin dust distribution, associated with the HI layer
   and older stellar populations, taken to represent the dust in the
   inter-arm regions.

\noindent
2. A optically thick dust distribution, associated with the molecular
   layer and with the young stellar populations, taken to represent
   the dust in the apiral arms.

\noindent
3. A clumpy component to represent Giant Molecular Cloud regions.

The model incorporates 3D radiative transfer, forward, and backward
scattering and realistic grain compositions and size
distributions. These are a significant step forward over the
simplistic slab models of the 90s and various studies have clearly
demonstrated the need for this level of sophistication when modeling
dust attenuation (\cite{witt}). The modeling of the
attenuation-inclination relation (Fig.~2) involves one free parameter,
the mean central face-on opacity, as all other aspects of the model
have been constrained via earlier multi-wavelenth observations of
nearby galaxies. The implied central face-on opacity is
$\tau_{B}^{f}=3.8\pm0.7$ which implies that the central region of disc
galaxies are, on average, optically thick. This will have serious
consequences for flux measurements of galaxy bulges.

\section*{Implication for galaxy transformation}
Based on the amount of dust we believe discs contain, Fig.~3 shows a
cartoon of a mid-type galaxy with a centrally optically thick disc as
it falls into a cluster. Simply by removing the dust, via some
unspecified mechanism, the galaxy is transformed to an earlier,
redder, and brighter system as the bulge is fully revealed and the
disc fades. While it is unlikely that this process explains the entire
transformation process from Sc to S0 it is undoubtedly part of the
story and may mitigate the need for dry mergers.

\begin{figure}[h]
\includegraphics[width=\textwidth]{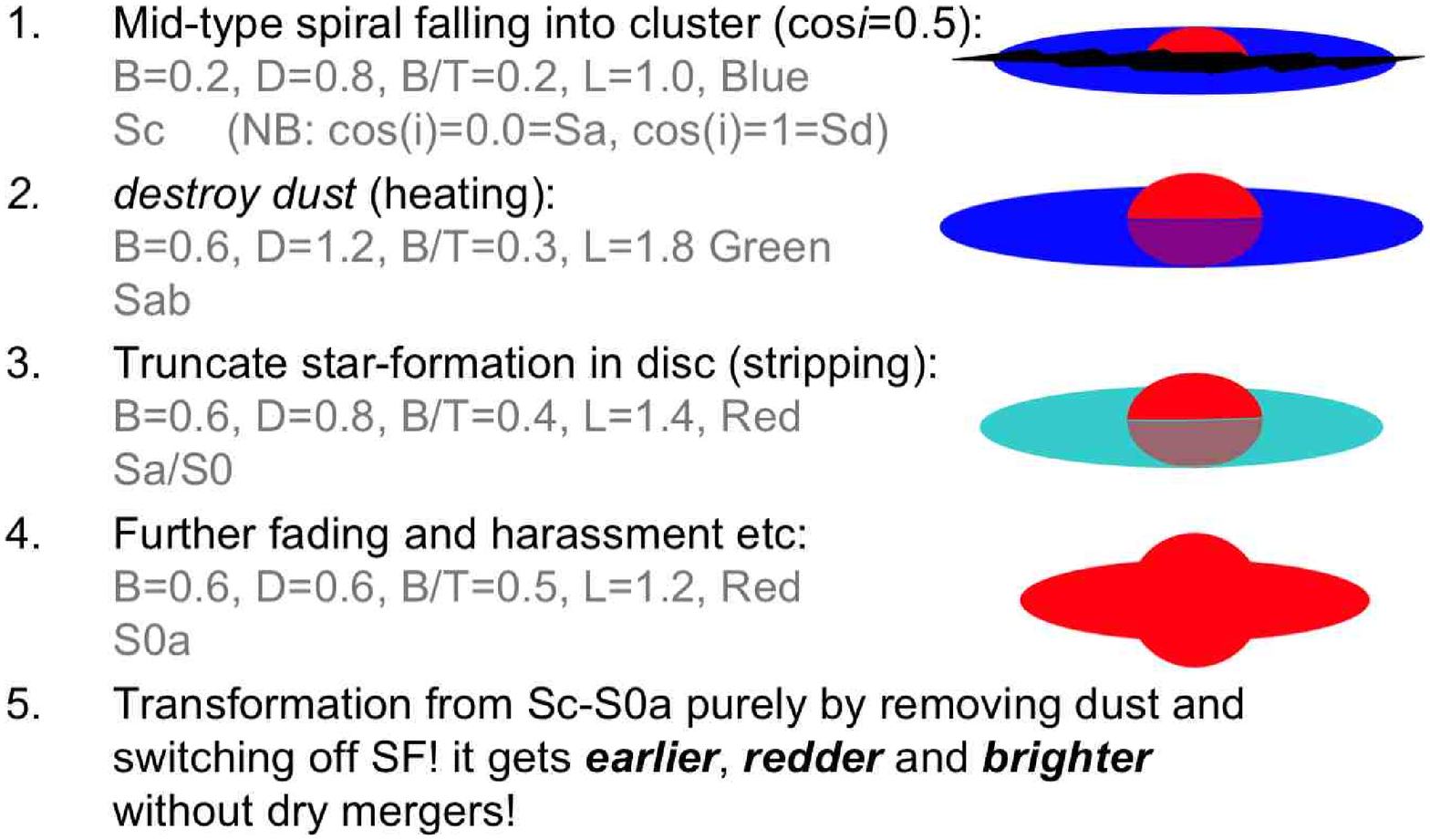}
\caption{A schematic showing the possible fate of a mid-type spiral entering a cluster. B refers to bulge fraction, D to disc fraction and L to luminosity.}
\end{figure}

\section*{The photon escape fraction and cosmic spectral energy distribution}

\vspace{-0.1cm}

In a forthcoming paper (Driver et al, in prep) we use our dust model
to derive the mean photon escape fraction versus wavelength (Fig.~4,
left) integrated over the nearby galaxy population and demonstrate
that the energy of starlight removed is consistent with the total
far-IR radiation (Fig.~4, right).

\begin{figure}
\includegraphics[width=6.7cm]{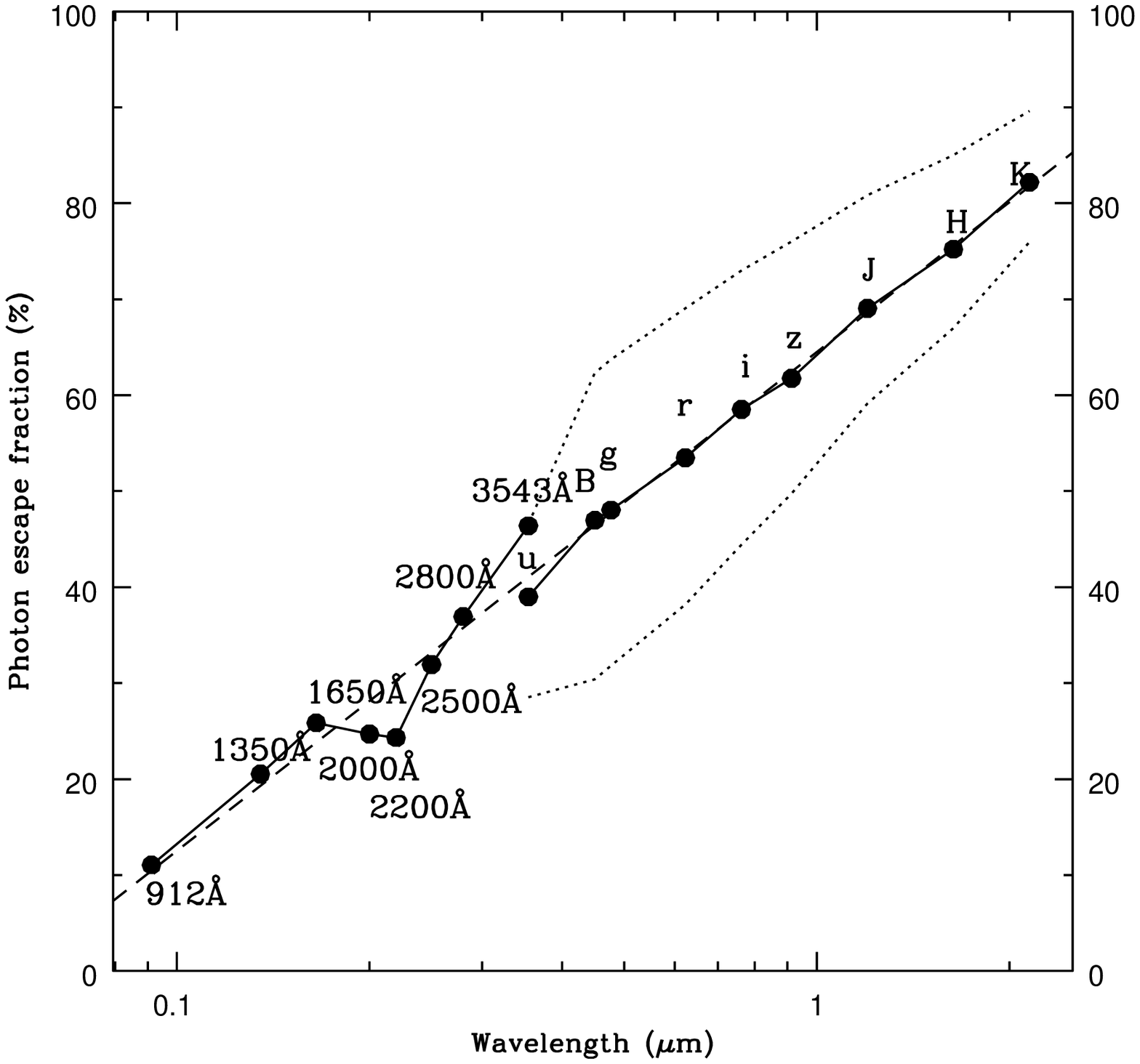}\includegraphics[width=6.0cm]{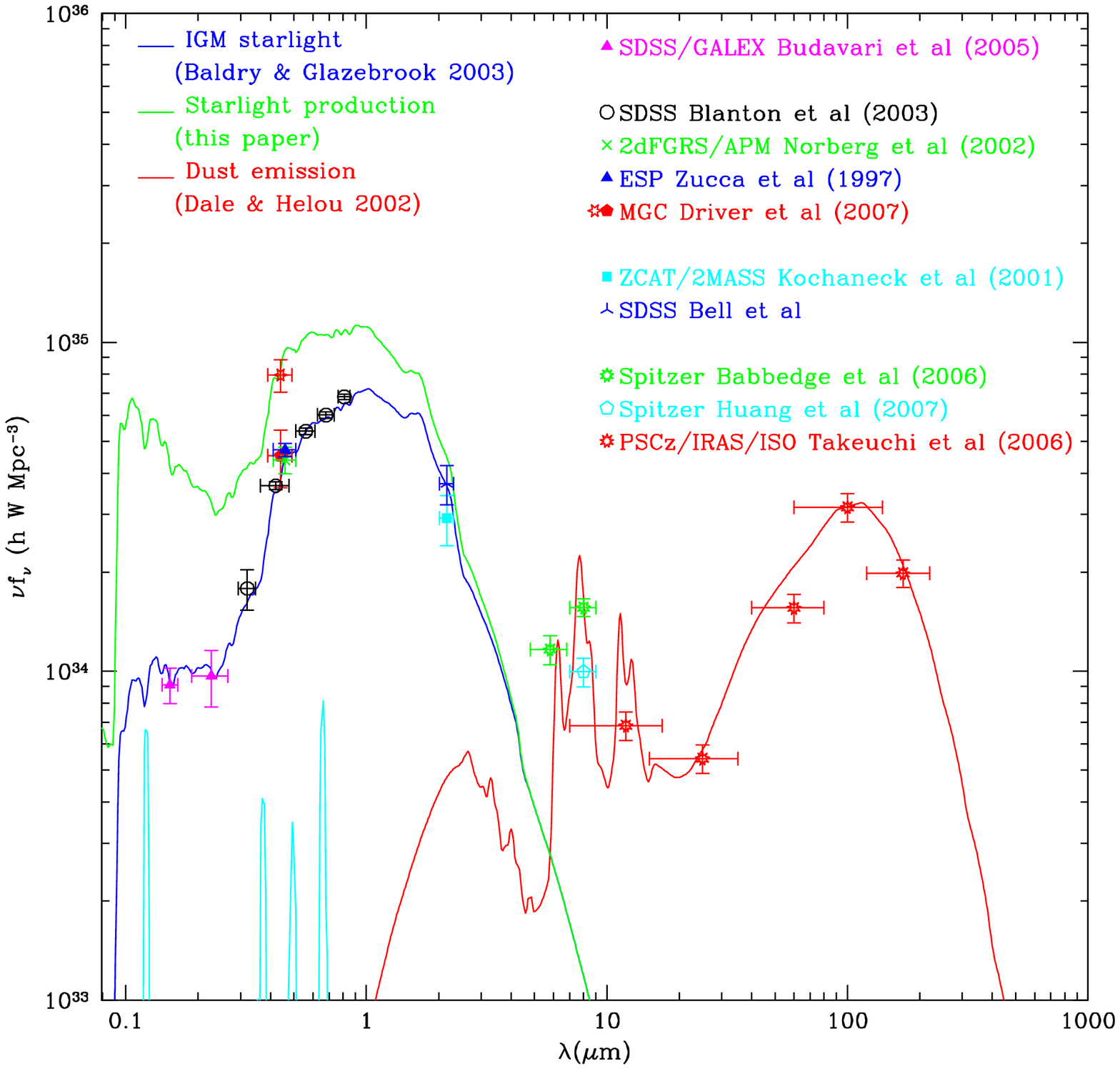}
\caption{(left) the fraction of photons which escape the galaxy population into the inter-galactic medium versus wavelength and (right) the cosmic spectral energy distribution showing the pre- (green), post- (blue attenuated starlight and the far-IR dust emission (red). Various data are shown as indicated.}
\vspace{-0.2cm}
\end{figure}

\vspace{-0.6cm}

\section*{Conclusions}

\vspace{-0.1cm}

\noindent
1. In the $B$-band galaxy discs are optically thick in their central regions.

\noindent
2. As a consequence total bulge flux can be attenuated by 0.8 to 2.5
   mags depending on inclination (and wavelength).

\noindent
3. Removing the dust from a mid- to late-type system, e.g., during cluster
   infall, will make a galaxy appear {\it significantly} earlier,
   redder, and brighter (in the $B$-band).

\noindent
4. Estimates of the luminosity density of galaxies in the nearby
   Universe significantly underestimate the total luminosity density
   produced by the integrated stellar population.

\noindent
5. The total energy of starlight absorbed equals the total far-IR
   emission (i.e., lost starlight = dust emission, 45\% of
   all starlight (energy) is attenuated by dust.

\end{document}